\documentclass[aps,prb,reprint,superscriptaddress]{revtex4-2}

\usepackage{graphicx}
\usepackage{amsmath}
\usepackage{braket}
\usepackage{xcolor}
\usepackage[normalem]{ulem}
\usepackage{notes2bib}
\bibnotesetup{
note-name = ,
use-sort-key = false
}
\newcommand{\expect}[1]{\langle#1\rangle}

\usepackage{hyperref}
\hypersetup{
    colorlinks=true,
    citecolor=teal,
    linkcolor=teal,
    filecolor=blue,      
    urlcolor=teal,
}
\usepackage[nameinlink,capitalise]{cleveref}

\usepackage{orcidlink}

\begin{document}

%Title of paper
\title{Effects of on-site Gaussian disorder in the ferromagnetic Blume-Capel model}

% List of authors 
\author{Kimberly J. Silverman}
\altaffiliation{These authors contributed equally to this work.}
\author{Daria Lhommedieu}
\altaffiliation{These authors contributed equally to this work.}
\author{Alykhan Rajan}
\author{Kathryn A. Cooper}
\affiliation{Department of Physics, Harvey Mudd College, 301 Platt Blvd, Claremont, CA 91711, USA}
\author{Richard T. Scalettar~\orcidlink{0000-0002-0521-3692}}
\affiliation{Department of Physics, University of California, Davis, CA 95616, USA}
\author{Eduardo Ibarra-Garc\'ia-Padilla~\orcidlink{0000-0001-9165-0444}}
\email[]{eibarragp@g.hmc.edu}
\affiliation{Department of Physics, Harvey Mudd College, 301 Platt Blvd, Claremont, CA 91711, USA}
\date{\today}

\begin{abstract}
We conduct a numerical study of a generalization of the Blume-Capel model in which the crystal field $D_i$, which controls the local vacancy density, is site dependent rather than uniform. We employ Markov Chain Monte Carlo to compute the shifted phase transition line $T_c$ as a function of the width $\sigma$ of a gaussian distribution $P(D_i) \sim \exp( -(D_i - \bar{D})^2/2 \sigma^2)$ at fixed $\bar{D}$. We characterize how the presence of this on-site Gaussian disorder reduces $T_c$ across the second order transition line, and discuss how our results suggest that $T_c$ remains finite even at large values of $\sigma$. We also determine the effects of disorder on the tricritical point of the conventional, uniform, Blume-Capel model, finding that it significantly suppresses the first-order phase transition, eliminating it entirely at large values of $\sigma$. 
\end{abstract}

\maketitle

\section{Introduction}

The Blume-Capel (BC) model~\cite{blume1966,CAPEL1966} generalizes the Ising model~\cite{Ising_review,EIGP2016,Onsager} to allow for ``vacant sites''.  This is accomplished by extending the values of the local spin degree of freedom from $s_{i} = \pm 1$ to $s_{i} = \pm 1, 0$.  The latter value, $s_{i}=0$, reflects the absence of a spin on site $i$.  These models have attracted continued interest in the statistical physics community owing to their rich phase diagram in the plane of temperature and crystal field $D$, which controls the $s_{i}=0$ density~\cite{Saul1974,lajzerowicz1975,Burkhardt1976,newman1983,Beale1986,Care_1993,Deserno1997,dillon2010,Wang2024,Rachadi2003}. In the limit where the crystal field is large and negative ($D\to-\infty)$, the Ising critical temperature is recovered. As the vacancy density increases, $T_{c}$ falls. Eventually, at sufficient diminishment of magnetic sites, long range order does not occur even at zero temperature and $T_{c}=0$. Along the route in which $T_{c}$ decreases there is a tricritical point and the original second order Ising transition becomes first order.

Although the BC model was originally introduced as a spin-1 description of magnetic phase transitions, its extension by Blume, Emery, and Griffiths (BEG) was subsequently proposed to understand phase separation and tricritical behavior in $^{3}$He--$^{4}$He mixtures \cite{Blume1971,Maritan1992}. These models have also found applications in explaining the behavior of ternary fluids~\cite{Mukamel1974,Gu1994Ternary,PenaLara2022}, phenomena of inverse melting~\cite{Schupper2005}, and semiconductor alloys~\cite{NiIwata1995}. The BEG model is more natural for alloy applications than the minimal BC model because it includes an extra term which provides an independent interaction depending on the species occupancy, allowing a mapping to the different pair energies in multicomponent alloys.

In  this manuscript, we consider the BC model with random crystal field $D_i$.   Our motivation is two-fold.  First, we would like to use unbiased Monte Carlo methods, rather than mean field approaches, to understand how first- and second-order transitions, as well as tricritical points, are modified in the presence of disorder. For example, the introduction of an infinitesimal amount of disorder in the bond strength $J_{ij}$ in the BEG model is known to change the nature of the first-order transition into a continuous transition or even destroy it altogether~\cite{Aizenman1989,Hui1989}. 
Second, random crystal fields are often invoked as part of the use of the BC and BEG models in alloy modeling~\cite{Buzano1994}, and a variation of the BC model with random exchange constants has been used to study critical interfacial adsorption phenomena~\cite{Fytas2019}. Our work generates precise numerical data concerning these potential applications.

Random-crystal-field versions of the Blume-Capel model have been investigated previously using a variety of analytical and approximate methods. Early mean-field studies considered, in particular, Gaussian-distributed crystal fields and established that quenched disorder can substantially modify the first-order and tricritical structure of the clean model~\cite{Carneiro1989,Carneiro1990}. Other mean-field, cluster-variation, and infinite-range treatments have explored discrete, most commonly bimodal, distributions of the crystal field and have found a variety of disorder-induced changes in the topology of the phase diagram, including the disappearance or replacement of tricritical behavior and the occurrence of reentrant and multicritical phenomena~\cite{Branco1997,Santos2015,Sumedha2020,Yessoufou2023}. A real-space renormalization group study of the two-dimensional model with bimodal random crystal fields found that randomness eliminates the first-order transition and leaves a line of continuous transitions~\cite{Branco1997}.

Numerical work on the finite-dimensional random-crystal-field model has been more limited. In particular, Vatansever et al.~performed high-precision Wang--Landau Monte Carlo simulations of the square-lattice Blume-Capel model with a bimodal distribution of the crystal field, focusing primarily on critical scaling and universality along the continuous-transition portion of the phase boundary and finding evidence for Ising critical behavior with logarithmic corrections; they also presented evidence that disorder rounds the clean first-order transition into a continuous one~\cite{Vatansever2020}.

In contrast, the present work considers a continuous Gaussian distribution of $D_i$ on the two-dimensional nearest-neighbor square lattice and follows the evolution of both the second-order transition temperature and the first-order/tricritical regime as the width $\sigma$ of this distribution is varied. Although Gaussian random crystal fields have been considered previously at the mean-field level~\cite{Carneiro1989,Carneiro1990}, to our knowledge their thermodynamic phase behavior in the two-dimensional short-range Blume-Capel model has not been characterized by Monte Carlo simulation.

The novelty of the present work lies in providing a direct Monte Carlo characterization of the two-dimensional short-range model with continuously distributed Gaussian disorder in the local crystal field, and in quantitatively comparing the robustness of its continuous transition with the rapid disorder-induced suppression of the first-order regime and tricritical point.

The paper is organized as follows. In Sec.~\ref{sec::Model_Methods} we present the model, details of the methods, and the observables computed. In Sec.~\ref{sec::Results} we present our results, focusing first on the effects of disorder on the second order transition, followed by the effects on the first order and the tricritical point. Finally, in Sec.~\ref{sec::Conc} we conclude.

\section{Model and methods}\label{sec::Model_Methods} 

\subsection{Model}
We study the Blume-Capel model in the square lattice in the presence of on-site disorder. The Hamiltonian is expressed as
\begin{equation}\label{eq:Hamilt}
H = -J \sum_{\expect{i,j}} s_i s_j +\sum_i D_i s_i^2,
\end{equation} 
where $s_i = \pm 1,0$ on site $i= 1,2,\dots,N$ in a two-dimensional square lattice with $N$ lattice sites, $\langle ij \rangle$ denotes the sum over nearest-neighbors on a square lattice, $J>0$ is the ferromagnetic coupling strength, and the crystal field $D_i$ is drawn from a gaussian distribution,
\begin{equation}
    P(D_i) = \frac{1}{\sigma \sqrt{2 \pi}}\exp\left[- \frac{(D_i-\bar{D})^2}{2\sigma^2}\right],
\label{eq:probofD}
\end{equation}
centered at $\bar{D}$ with standard deviation $\sigma$.  We set the energy scale $J=1$.

In the absence of disorder ($\sigma=0$), i.e. when the crystal field is uniform, there are two relevant limits: As $\bar{D}/J \to -\infty$, the system approaches the Ising limit, and $T_c/J \to 2.269$. Meanwhile, for $\bar{D}/J>2$, $T_c=0$ since the cost to occupy pairs of sites exceeds the energy lowering from $J$.

\subsection{Monte Carlo and observables}

We perform classical Monte Carlo simulations in square lattices with linear sizes up to $L=64$. We find the simple single spin flip approach using the Metropolis algorithm to accept or reject proposed changes to $s_i$ provides sufficient accuracy without recourse to more sophisticated cluster~\cite{wolff1989,ferrenberg1992,PhysRevB.54.359}, parallel tempering~\cite{Katzgraber2006}, or Wang-Landau approaches~\cite{wang2001}. Near the first order transition, where coexistence of phases is expected, we conduct simulations considering fully polarized systems (forward runs) and then considering fully depleted lattices (backward runs) to diagnose hysteretic behavior. In order to circumvent non-ergodic sampling in the backward runs, we introduce small external magnetic field $h$ to perform annealing during the warmup sweeps, which is then turned off during the measurement sweeps. For all calculations, we consider a minimum of $5\times10^3$ warmup sweeps, and $1\times10^{5}$ measurement sweeps.  We calculate the thermal averages of observables over at least 50 realizations of disorder at each value of $\sigma/J$. In Appendix~\ref{App::A} we present evidence that the results for these linear sizes and number of realizations of disorder are already self-averaged and provide converged results. Results in the main text are presented for $L=64$ unless otherwise specified. 

We compute the energy $E = \expect{H}/N$, the density $n = \sum_i\expect{ s_i^2}/N$, the magnetization $M = \sum_i\expect{s_i}/N$, the specific heat $ C = \beta^2\left[\langle E^2 \rangle - \langle E \rangle ^2 \right]$, the magnetic susceptibility $ \chi = \beta\left[\langle M^2 \rangle - \langle M \rangle ^2 \right]$, the compressibility $ \kappa = \beta\left[\langle n^2 \rangle - \langle n \rangle ^2 \right]$, and the Binder ratio $U = 1 - (1/3) \langle M^4 \rangle/\langle M^2 \rangle ^2$. Near the tricritical point, we also compute the area of the hysteresis loop $A_n = \oint n d\bar{D}/J$.

These observables provide valuable knowledge about the physics: The density and the compressibility measure the effects of the crystal field and the disorder strength on the filling fraction, the magnetization is a useful order parameter for the paramagnetic-ferromagnetic transition, the specific heat and the magnetic susceptibility provide information about the temperature scales at which the degrees of freedom cease to fluctuate, the Binder ratio is a powerful tool to locate phase transitions, and the hysteresis loop area directly corresponds to the thermodynamic work required to drive the material between coexisting phases.

\subsection{Local density analysis}

It is useful to compare our numerical results against simple approximations, which provide insight into the physics. Of interest to our work, in studies of confined atomic cases with a smoothly varying (confining) chemical potential $\mu_i$, a commonly used approximation, named \textit{local density approximation} (LDA), relies on modeling the local behavior for a particular $\mu_i$ as that of the homogeneous lattice at that parameter value. In the same spirit, a crude picture of the effect of $\sigma/J$ on $T_{c}/J$ is,
\begin{equation}
T_c(\bar{D},\sigma) = \int dD_i \, P(D_i) \, T_{c,{\rm unif}}(D_i)
\label{eq:TC_LDA}.
\end{equation}
where $P(D_i)$ is the Gaussian distribution of Eq.~\eqref{eq:probofD} which is parameterized by $\bar{D}$ and $\sigma$ and $T_{c,\rm{unif}}$ is the critical temperature of the uniform system.

\section{Results}\label{sec::Results}

In this section we present our results of the Blume-Capel model in the presence on an on-site gaussian disorder. We first discuss the effects of the disorder strength $\sigma/J$ across the second order transition line at fixed $\bar{D}/J$ in Section~\ref{sec::2nd}. Then we examine its effects across the first order transition line at fixed $T/J$ in Section~\ref{sec::1st}. 

\subsection{Effects of $\sigma/J$ on the second order transition}\label{sec::2nd}

\begin{figure}[tbp!]
    \centering
    \includegraphics[width=\linewidth]{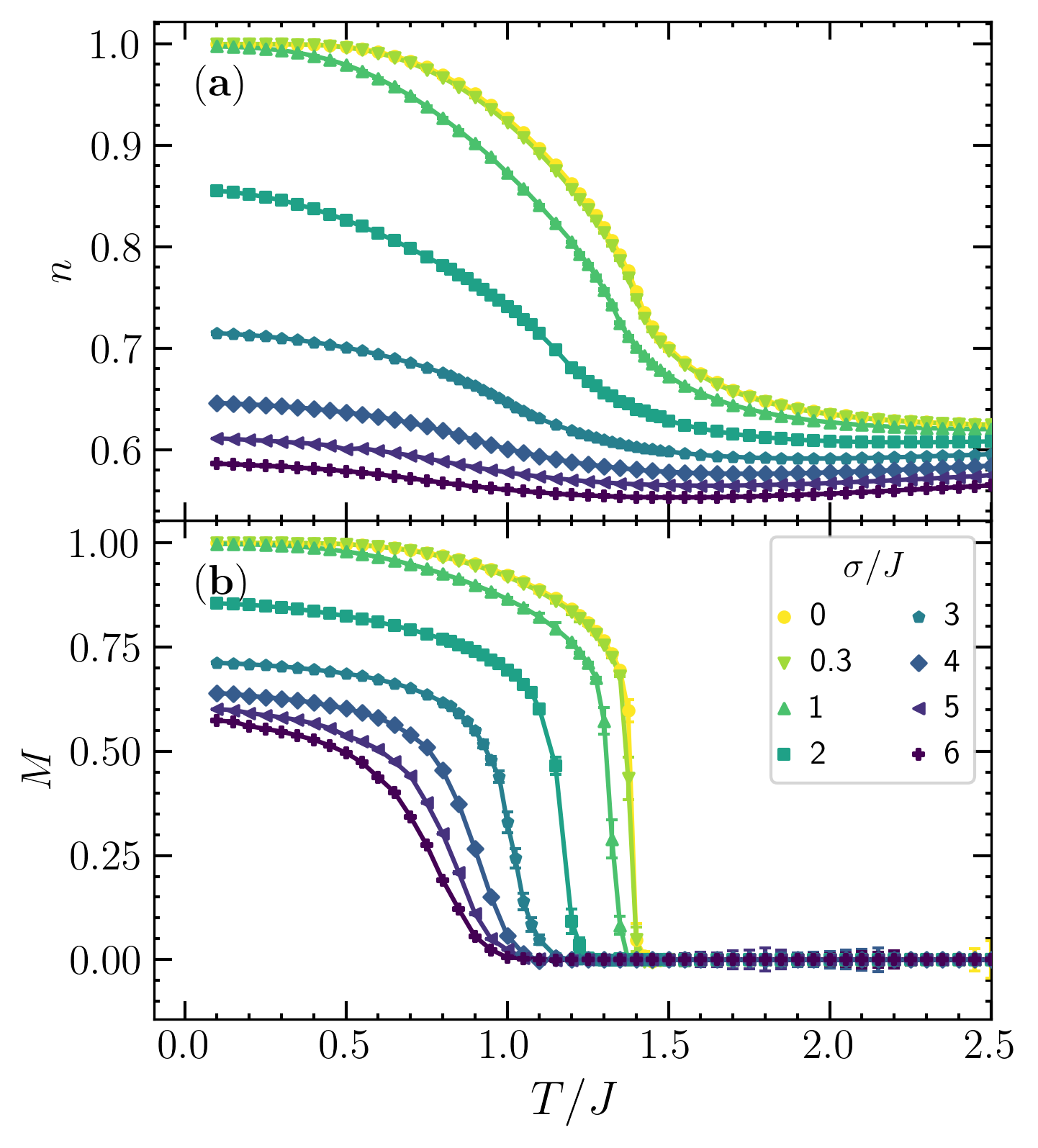}
    \caption{(a) Density $n$ and (b) magnetization $M$ as functions of $T/J$ at $\bar{D}/J =1$ for different $\sigma/J$.
    } 
    \label{fig:densmag}
\end{figure}

Fig.~\ref{fig:densmag}(a) shows the dependence of density $n$ on the temperature for different amounts of disorder at fixed $\bar{D}/J =1$. At high temperatures, where all spin states are equally probable, $n=2/3$. As the temperature is lowered, in the absence of disorder ($\sigma=0$), $n$ increases monotonically to 1, indicating a fully occupied lattice. Increasing $\sigma/J$ up to 0.3, there is no appreciable change in the density. As $\sigma/J$ further increases past $\sigma/J=1$, the density decreases and the ground state is no longer fully occupied. At such large values of $\sigma/J$, $n$ is no longer monotonic with $T/J$, and for the largest values of $\sigma/J=5,6$, the density remains close to its high temperature limit $n\sim2/3$. 

The magnetization results presented in Fig.~\ref{fig:densmag}(b) show that despite a partially filled lattice when disorder is present, there is still ferromagnetic order. This is supported by the fact that $M\approx n$ as $T/J\to0$, highlighting that as the lattice is filled, is energetically favorable to do so ferromagnetically. However, as $\sigma/J$ increases, two main effects are visible. First, the magnetization upturn occurs at lower temperatures. And second, the slope at which $M$ increases as $T/J$ is lowered is smaller. For $\sigma/J \leq 0.3$, $M=0$ above $T_c/J$ and grows rapidly as the temperature is lowered across the transition. As $\sigma/J$ increases, this upturn becomes smoother.

\begin{figure}[tbp!]
    \centering
\includegraphics[width=\linewidth]{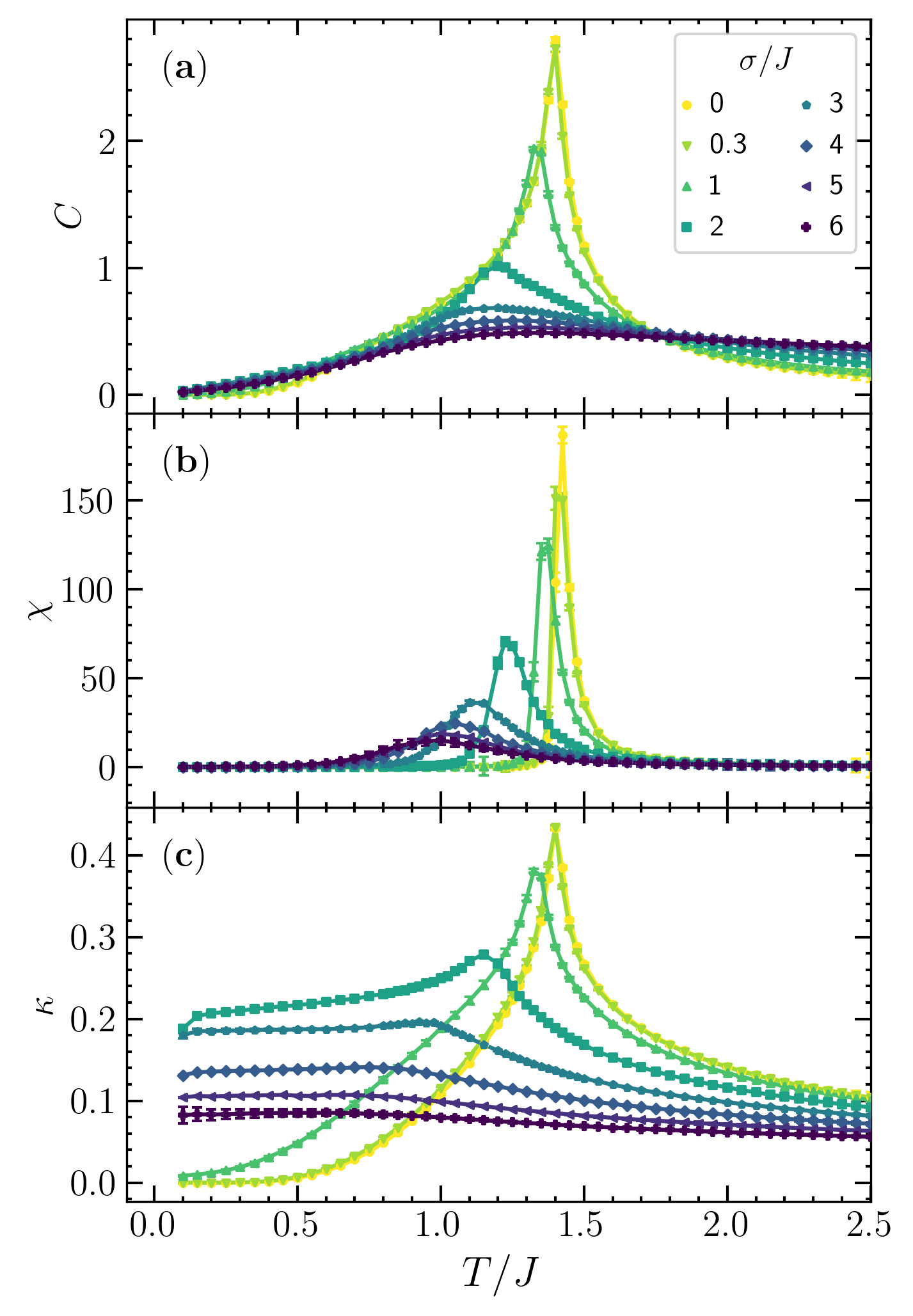}
    \caption{(a) Specific heat $C$, (b) magnetic susceptibility $\chi$, and (c) compressibility $\kappa$ as functions of $T/J$ at $\bar{D}/J =1$ for different $\sigma/J$.} 
    \label{fig:CXK}
\end{figure}

The transition temperature $T_c/J$ of the paramagnetic-ferromagnetic transition moves to lower temperatures with increasing disorder as evidenced by the specific heat $C$ in Fig.~\ref{fig:CXK}(a). This shift to lower temperatures is accompanied by a broadening of the peaks as $\sigma/J$ increases. Such behavior can be understood in terms of the relationship between the entropy $S$ and the specific heat $C$. Since $dS = (C/T)dT$,
\begin{equation}
    S_\infty = \int_{0}^\infty \frac{C}{T'}dT'.
\end{equation}
For the Blume-Capel model, $S_\infty = \ln 3$. Because this quantity is constant, to satisfy the sum rule, the peaks in $C$ must also broaden as they move to lower $T/J$ as $\sigma/J$ increases. To extract the precise $T_c/J$, we analyze the magnetic susceptibility $\chi$, since this observable displays well resolved peaks even at the largest $\sigma/J$ considered [see Fig.~\ref{fig:CXK}(b)].

In Fig.~\ref{fig:CXK}(c) we present the temperature and disorder dependence of the compressibility $\kappa$. For $\sigma/J \leq 0.3$, $\kappa \to 0$ in the ground state, indicating a completely filled lattice. However for $\sigma\geq1$, $\kappa$ remains finite, showing that the lattice is only partially filled. This can be explained by the presence of large values of $D_i$ in the crystal field distribution that ensure that vacancies remain present even at low temperatures. In other words, the system is still compressible, but it would be energetically expensive to occupy these sites.

\begin{figure}[tbp!]
    \centering
\includegraphics[width=\linewidth]{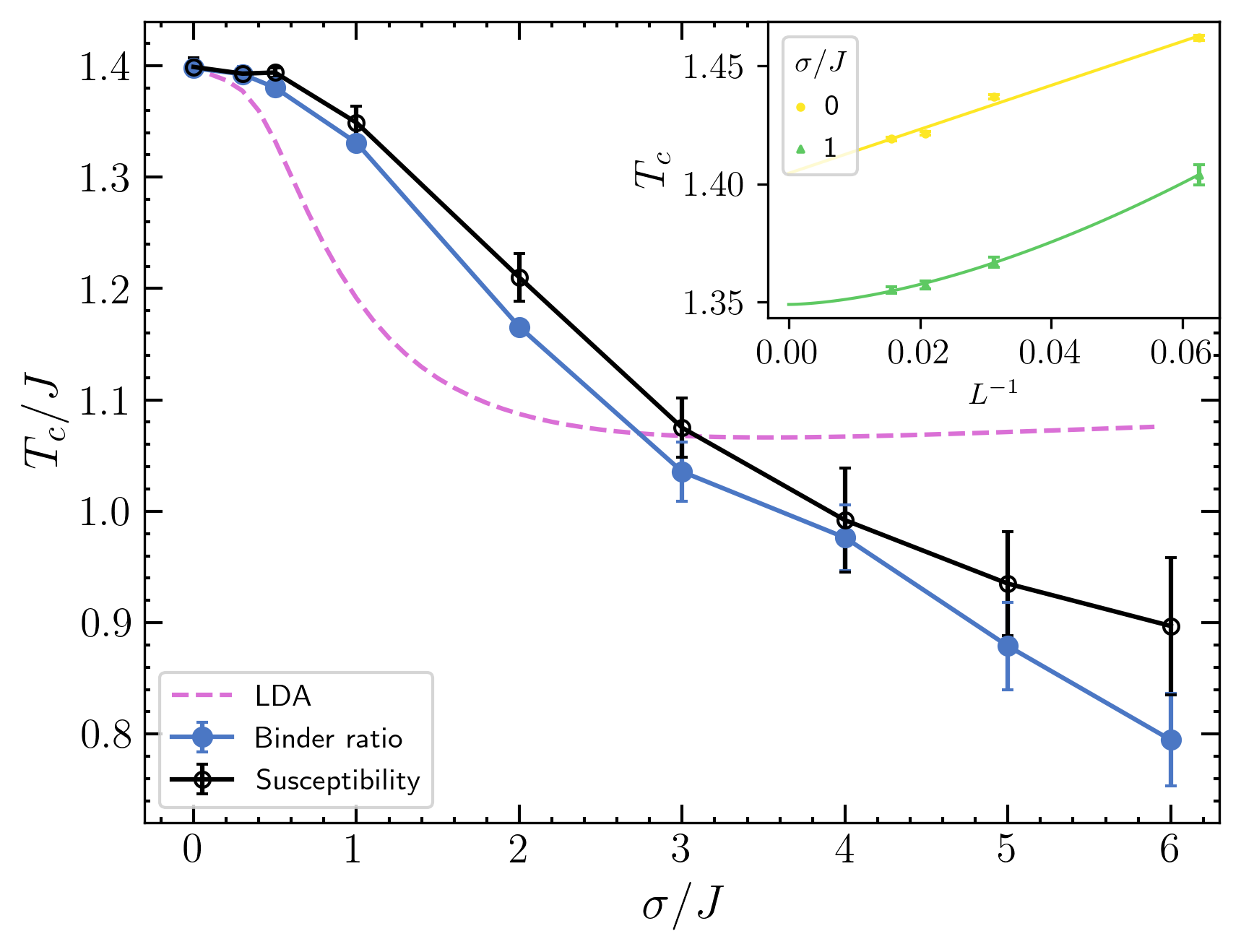}
    \caption{$T_c/J$ as a function of $\sigma/J$ at $\bar{D}/J=1$. Blue solid markers: $T_c/J$ extracted from the crossings in the Binder ratio, where error bars are estimated as the difference between the predicted $T_c/J$ from the crossings between $L=64,48$ and $L=64,32$ (see Fig.~\ref{fig:Binder_example} for an example). Black open markers: $T_c/J$ from the finite-size scaling of the peak locations in $\chi$, with the error bars given from the fit parameters errors. The inset illustrates the finite-size scaling procedure for $\sigma/J=0,1$, where we fit to a power law $T_c(L) = T_c(\infty) + bL^{-\nu}$. Pink dashed line: $T_c/J$ from the LDA analysis given by Eq.~\eqref{eq:TC_LDA}.
    } 
    \label{fig:TvsS}
\end{figure}

Fig.~\ref{fig:TvsS} shows how the second order transition gets modified in the presence of disorder. As $\sigma/J$ increases, $T_c/J$ decreases, i.e. $dT_c/d\sigma <0$. However, even at $\sigma/J=6$ $T_c/J$ remains finite and is still surprisingly large. The precise $T_c/J$ are obtained via two complementary observables: the Binder ratio and the finite-size scaled susceptibility (see inset for examples of the finite-size scaling for $\sigma/J=0,1$). These agree with each other within error bars for most values of $\sigma/J$, reinforcing our findings. In addition to the numerical results, we also present the predicted $T_c/J$ using the LDA. While this approximation is unable to fully capture the behavior observed in the Monte Carlo results, it correctly reproduces the disorder dependence at $\sigma/J\ll 1$ and predicts that $T_c$ remains finite in the limit  $\sigma/J \to \infty$~\footnote{Concretely, within the LDA, $T_c\to T_c^\mathrm{Ising}/2$ as $\sigma/J \to \infty$ because it averages over large regions where $T_c=0$ ($\bar{D}/J >2$) and regions where $T_c$ approaches the Ising limit ($\bar{D}/J \to -\infty$).}, further supporting the conclusion that $T_c/J$ remains finite at large values of $\sigma/J$.  Calculations at $\bar{D}/J=0$ also show a similar trend with $\sigma/J$ (not shown).

\subsection{Effects of $\sigma/J$ on the first order transition}\label{sec::1st}

\begin{figure}[tbp!]
\centering
\includegraphics[width=\linewidth]{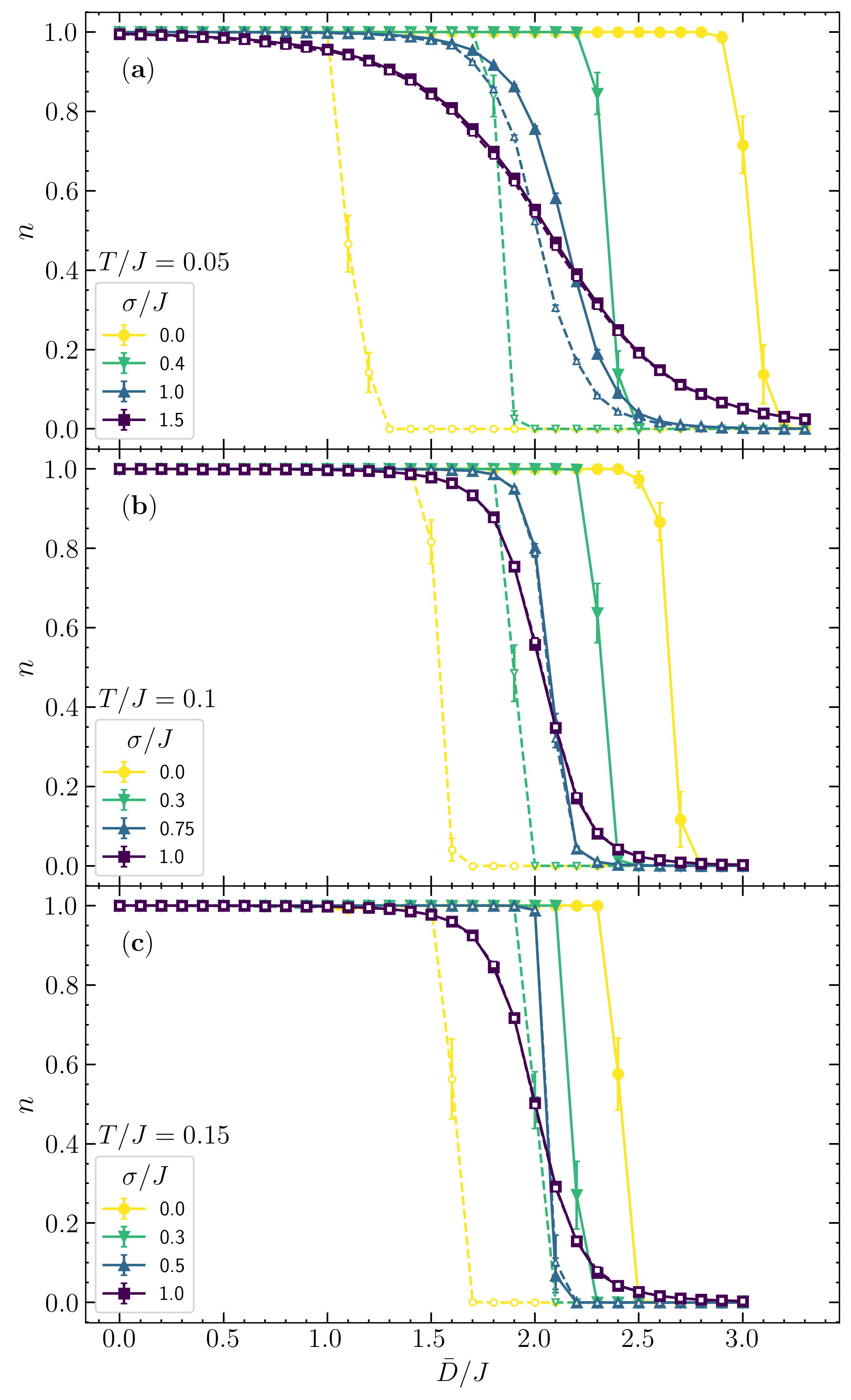}
    \caption{Density $n$ as a function of $\bar{D}/J$ for selected values of $\sigma/J$ at $T/J=0.05$, $ 0.1$, and $0.15$. Filled markers with solid lines indicate forward runs, and open markers with  dotted lines indicate backward runs. Backward runs employ a bias field $h\leq\sigma/4$ during warmup sweeps to perform annealing and ensure convergence.}
    \label{fig:hysteresis}
\end{figure}

Fig.~\ref{fig:hysteresis} presents the dependence of density $n$ on $\bar{D}/J$ for different $\sigma/J$. Results are displayed at three $T/J$ below the $\sigma=0$ tricritical point temperature $T_\mathrm{tricritical}/J = 0.6080(1)$~\cite{Kwak2015,Fytas2019,Zierenberg2017}. At the lowest temperature, $T/J=0.05$ [see Fig.~\ref{fig:hysteresis}(a)], the system can be considered to be in the ground state. In the absence of disorder ($\sigma=0$), $n$ displays a wide and sharp hysteresis loop centered around $\bar{D}/J\approx2$. The presence of such hysteretic behavior is a strong indicator of a first-order phase transition, where the loop area directly corresponds to the thermodynamic work required to drive the material between coexisting phases, in this case an empty lattice and the ferromagnetic state. As $\sigma/J$ increases, for $\sigma/J=0.4$, the area of the hysteresis loop decreases but it still remains sharp. For $\sigma/J=1.0$, the $n$ curve becomes smoother, but retains its hysteretic behavior. Finally, as disorder increases to $\sigma/J=1.5$ the loop vanishes and the first order transition disappears. Similar trends are observed at higher temperatures $T/J=0.1, 0.15$, where the initial $\sigma=0$ hysteresis loop shrinks with increasing temperature and vanishes at smaller values of $\sigma/J$ [see Figs.~\ref{fig:hysteresis}(b)-(c)]. 

\begin{figure}[tbp!]
    \centering
    \includegraphics[width=\linewidth]{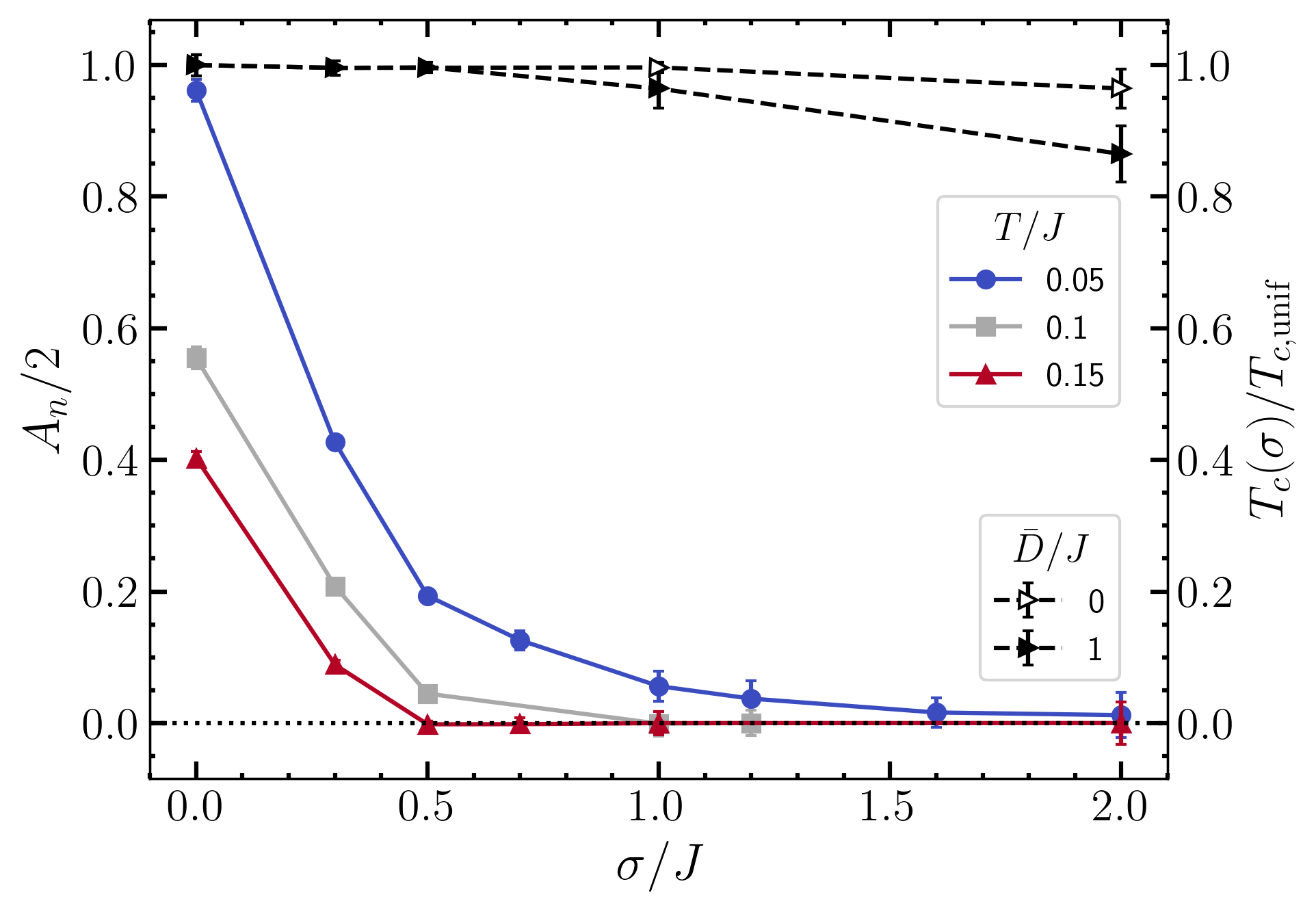}
    \caption{Left axis: Half of the hysteresis loop area $A_n/2$ as a function of $\sigma/J$ at selected temperatures $T/J$. Error bars are obtained by error propagation when calculating the area. Right axis: Rescaled second order transition $T_c (\sigma)/T_{c,\text{unif}}$ as a function of $\sigma/J$ for $\bar{D}/J =0,1$. }
    \label{fig:loop}
\end{figure}

Our results show that disorder eliminates the first order transition, consistent with the literature for other models of disorder~\cite{Aizenman1989,Hui1989}. This strong suppression of the first order transition is evidenced in Fig.~\ref{fig:loop} where we present $A_n/2$ as a function of $\sigma/J$ for $T <T_\mathrm{tricritical}$. Specifically, the area of the hysteresis loop decays rapidly as $\sigma/J$ increases, and vanishes at lower disorder strengths as temperature increases. This rapid suppression of the first order transition contrasts with the results observed along the second order phase transition: while $A_n/2$ decreases exponentially with $\sigma/J$, the decay in $T_c(\sigma)/T_{c,\mathrm{unif}}$ is much slower. As an example, a disorder strength of $\sigma/J=6$ is required  to suppress $T_c(\sigma)/T_{c,\mathrm{unif}}$ by a factor of 2. On the contrary, a small amount of disorder, $\sigma/J=0.4$, is enough to reduce $A_n$ by a factor of 4. 

\begin{figure}[tbp!]
\centering
\includegraphics[width=\linewidth]{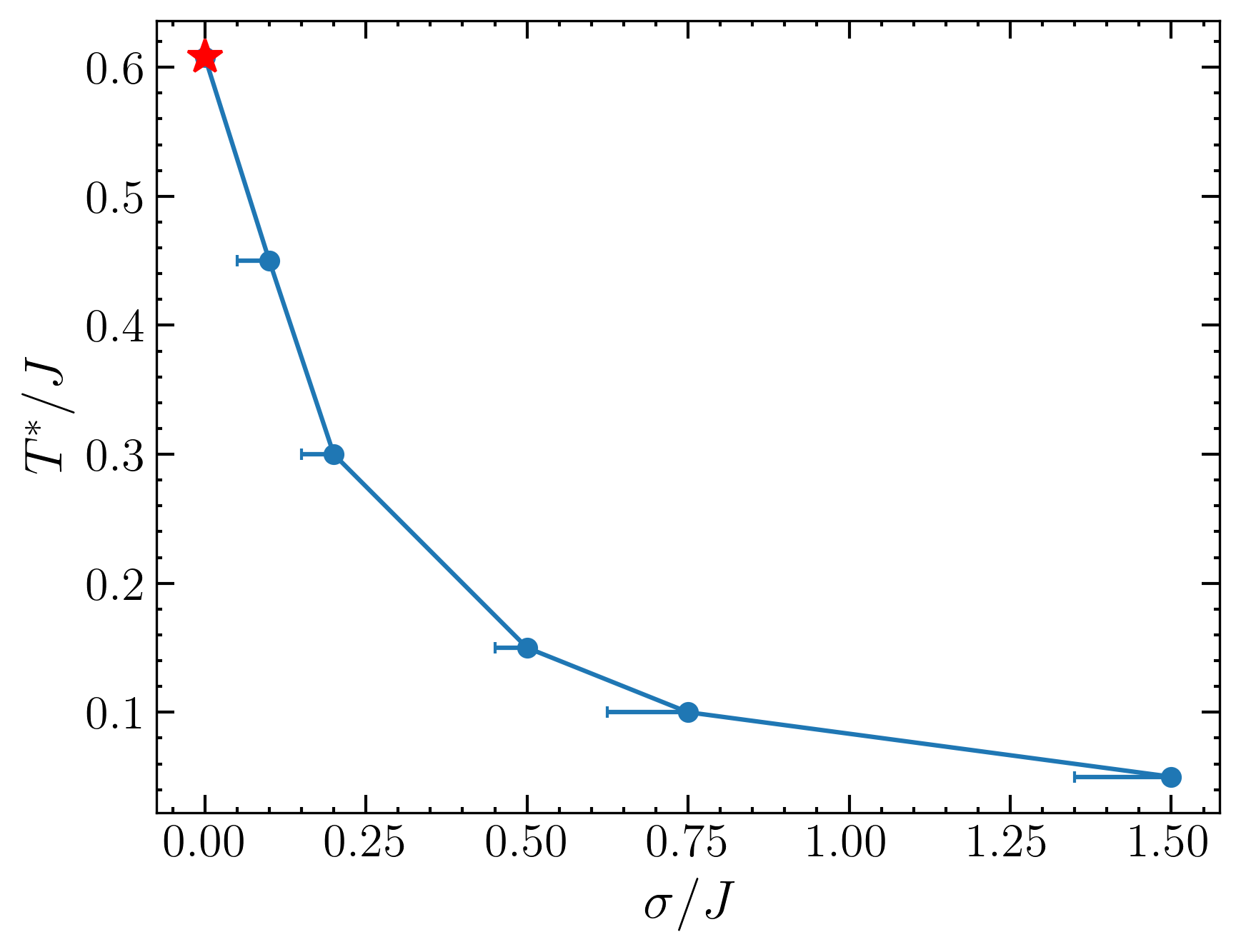}
    \caption{$T^*/J$, defined as the temperature at which the hysteresis loop closes, i.e. $A_n=0$ (within error bars), as a function of $\sigma/J$. Error bars provide a lower bound, and correspond to half the distance to the next smallest value of $\sigma/J$ for which $A_n>0$. The red star corresponds to the tricritical point temperature in the absence of disorder~\cite{Kwak2015}.}
    \label{fig:Tc}
\end{figure}

To further characterize the suppression of the first order transition, we define $T^*$ as the temperature at which the hysteresis loop closes at fixed $\sigma/J$. This quantity is presented vs $\sigma/J$ in Fig.~\ref{fig:Tc} and illustrates that the tricritical point temperature $T_\mathrm{tricritical}/J$ rapidly approaches zero as disorder increases.

\section{Conclusions}\label{sec::Conc}

In this work, we performed a Monte Carlo study of a generalization of the ferromagnetic Blume-Capel model to include on-site Gaussian disorder. We examined the dependence of thermodynamic observables and susceptibilities on $\sigma/J$ across the paramagnetic-ferromagnetic transition. We demonstrated that the introduction of disorder reduces $T_c/J$ along the second-order transition, but we predict that $T_c/J$ remains finite even at large values of $\sigma/J$. Furthermore, we showed that introduction of disorder strongly suppresses the first order transition, with the tricritical point temperature rapidly approaching zero as $\sigma/J$ increases. 
The results in this manuscript provide valuable fundamental insight into how first and second order transitions, and the
associated tricritical behavior, get modified in the presence of on-site Gaussian disorder. 

A particularly direct physical setting combining the two ingredients considered here is provided by $^{3}$He--$^{4}$He mixtures confined in porous media. While helium mixtures provided one of the original motivations for spin-1 Blume-Emery-Griffiths models, experiments on mixtures confined in silica aerogel have investigated how quenched spatial disorder modifies both phase separation and superfluidity.  Aerogel was found to detach the $^{3}$He--$^{4}$He coexistence region from the superfluid transition line, eliminating the bulk tricritical point and producing a miscible superfluid region rich in $^{3}$He~\cite{MaKimHrubeshChan1993,MaKimChan1994,ChanMuldersReppy1996}. 
Theoretical work aimed at the experimental situation of helium mixtures in an aerogel geometry using renormalization-group methods, 
found suppression of the tricritical temperature and new phase-separation structure~\cite{LopatnikovaBerker1997}.

Although the present Blume-Capel model does not contain the continuous superfluid degree of freedom of the helium mixture, a spatially random crystal field provides a simple classical setting in which to examine how quenched disorder modifies the phase-separation and tricritical physics that motivated these models and experiments.
Indeed, there is a qualitative parallel to our numerical results: the aerogel experiments found that disorder/confinement removes the ordinary tricritical meeting of the phase-separation and superfluid lines, while the present Monte Carlo simulations find rapid suppression of the clean BC first-order regime and tricritical temperature. 

\begin{acknowledgments}
This material is based upon work supported by the National Science Foundation under Award No. 2408259.
RTS is supported by the grant DOE DE-SC0014671 funded by
the U.S. Department of Energy, Office of Science.
\end{acknowledgments}

\appendix

\section{Self-averaging, finite-size effects, and the Binder ratio}\label{App::A}

\begin{figure}[tbp!]
    \centering
    \includegraphics[width=\linewidth]{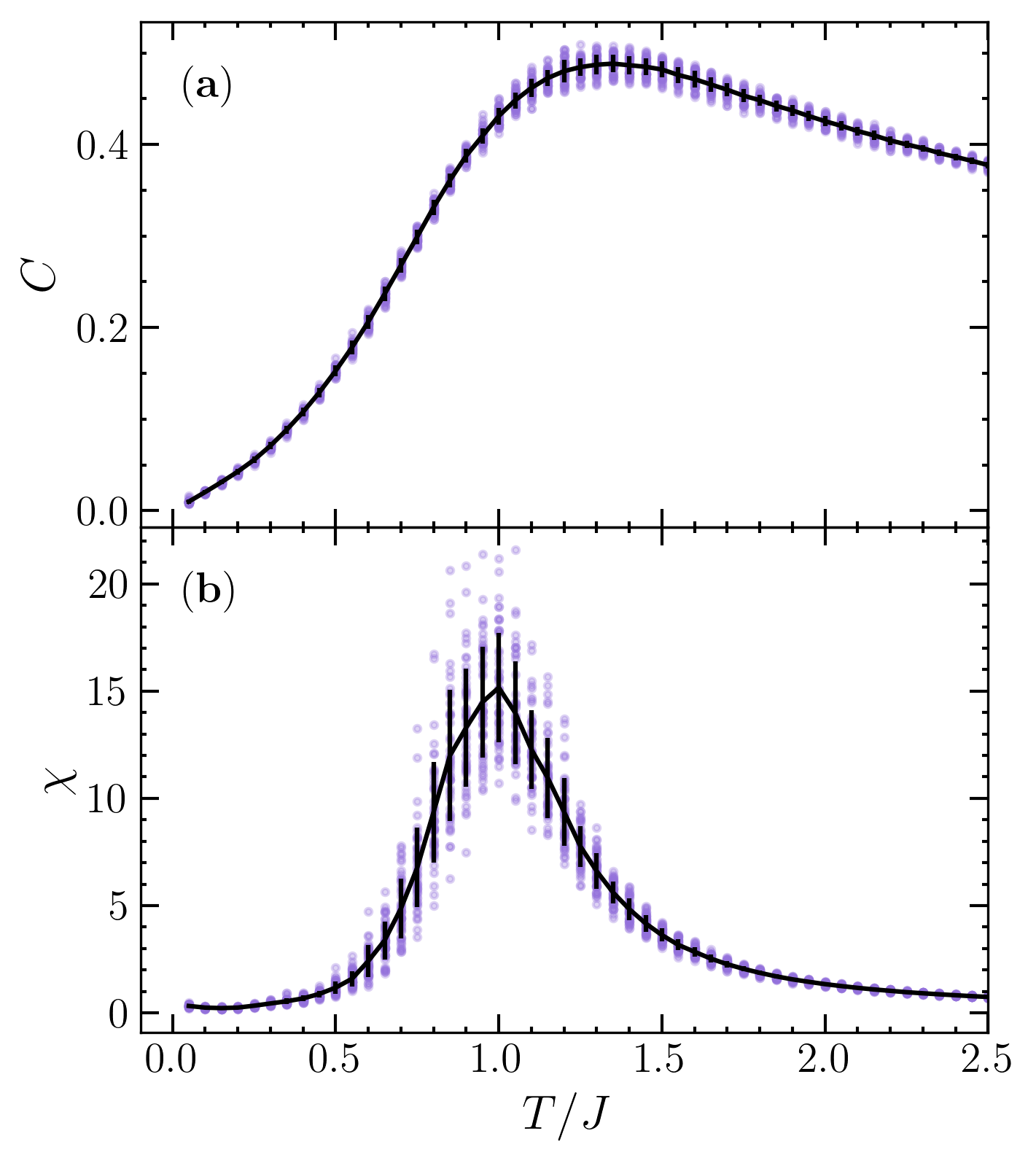}
    \caption{(a) Specific heat $C$ and (b) magnetic susceptibility $\chi$ for 50 realizations of disorder at $\bar{D}/J=1$ and $\sigma/J=6$ on a $L=64$ lattice. Purple markers correspond to different realizations of disorder, and the black line their mean. Error bars are the S.E.M. across all  realizations.}
    \label{fig:self_averaging}
\end{figure}

We investigate the convergence of our results with system size and number of realizations of disorder. First, we demonstrate in Fig.~\ref{fig:self_averaging} that for 50 realizations of disorder, the specific heat $C$ and the magnetic susceptibility $\chi$ display self-averaging, with a universal shift in the critical temperature across all realizations of disorder. These results validate the convergence of our results, and are presented at the largest amount of disorder ($\sigma/J=6$) on the largest system size ($L=64$) considered at $\bar{D}/J=1$ as functions of $T/J$. 

\begin{figure}[tbp]
\centering
\includegraphics[width=\linewidth]{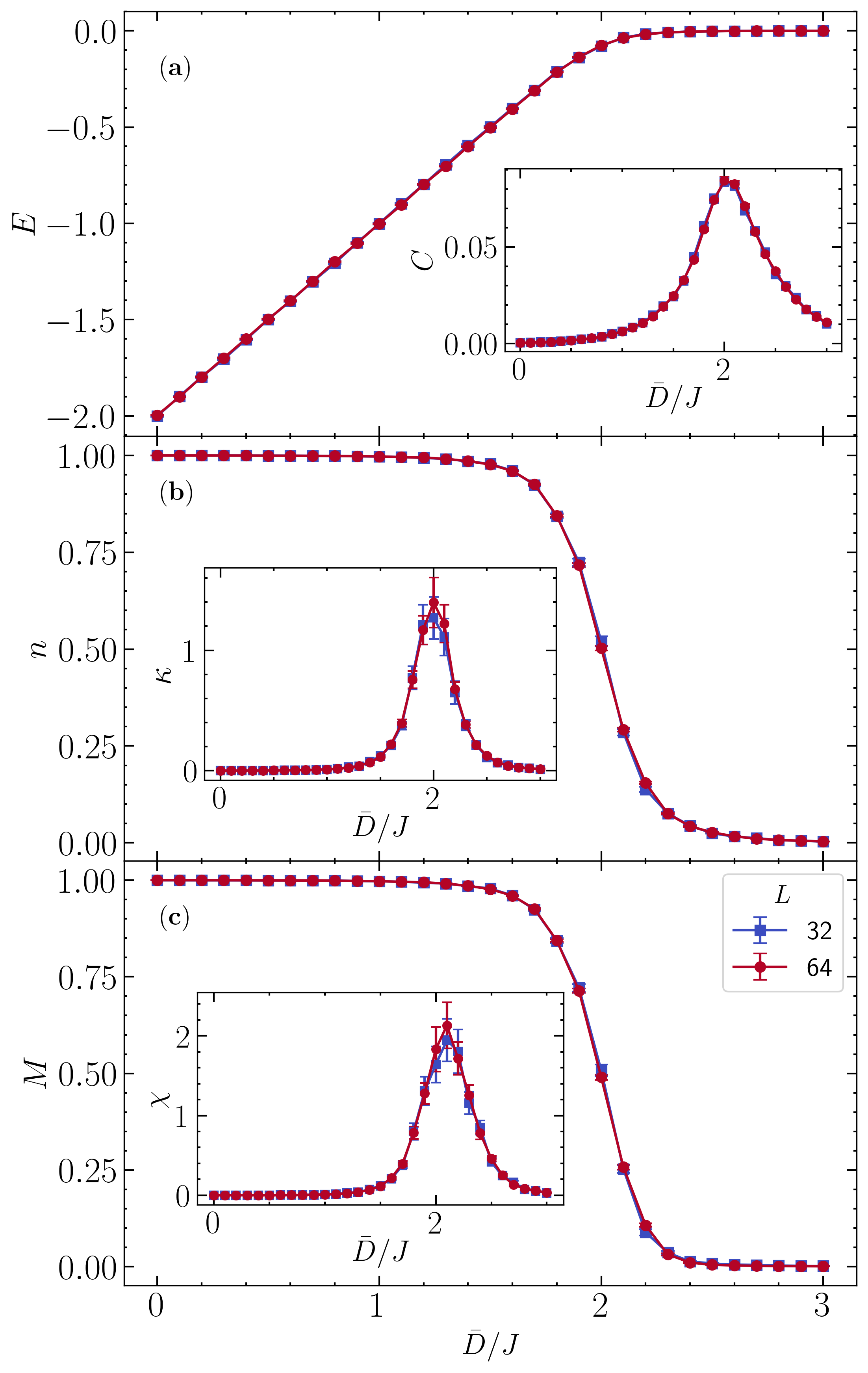}
    \caption{Energy $E$, density $n$, magnetization $M$, specific heat $C$, compressibility $\kappa$, and magnetic susceptibility $\chi$ as a function of $\bar{D}/J$ for $T/J=0.15$, $\sigma/J=1$, and $L=32,64$.}
    \label{fig:L_effects}
\end{figure}

Second, we show in Fig.~\ref{fig:L_effects} the negligible finite-size effects on thermodynamic observables, namely $E$, $n$, $M$, and their associated susceptibilities $C$, $\kappa$, and $\chi$ as functions of $\bar{D}/J$. Results are presented at $T/J =0.15$ for $L=32,64$ at $\sigma/J=1$, further supporting the strong suppression of the first-order transition.

\begin{figure}[bp!]
    \centering
    \includegraphics[width=\linewidth]{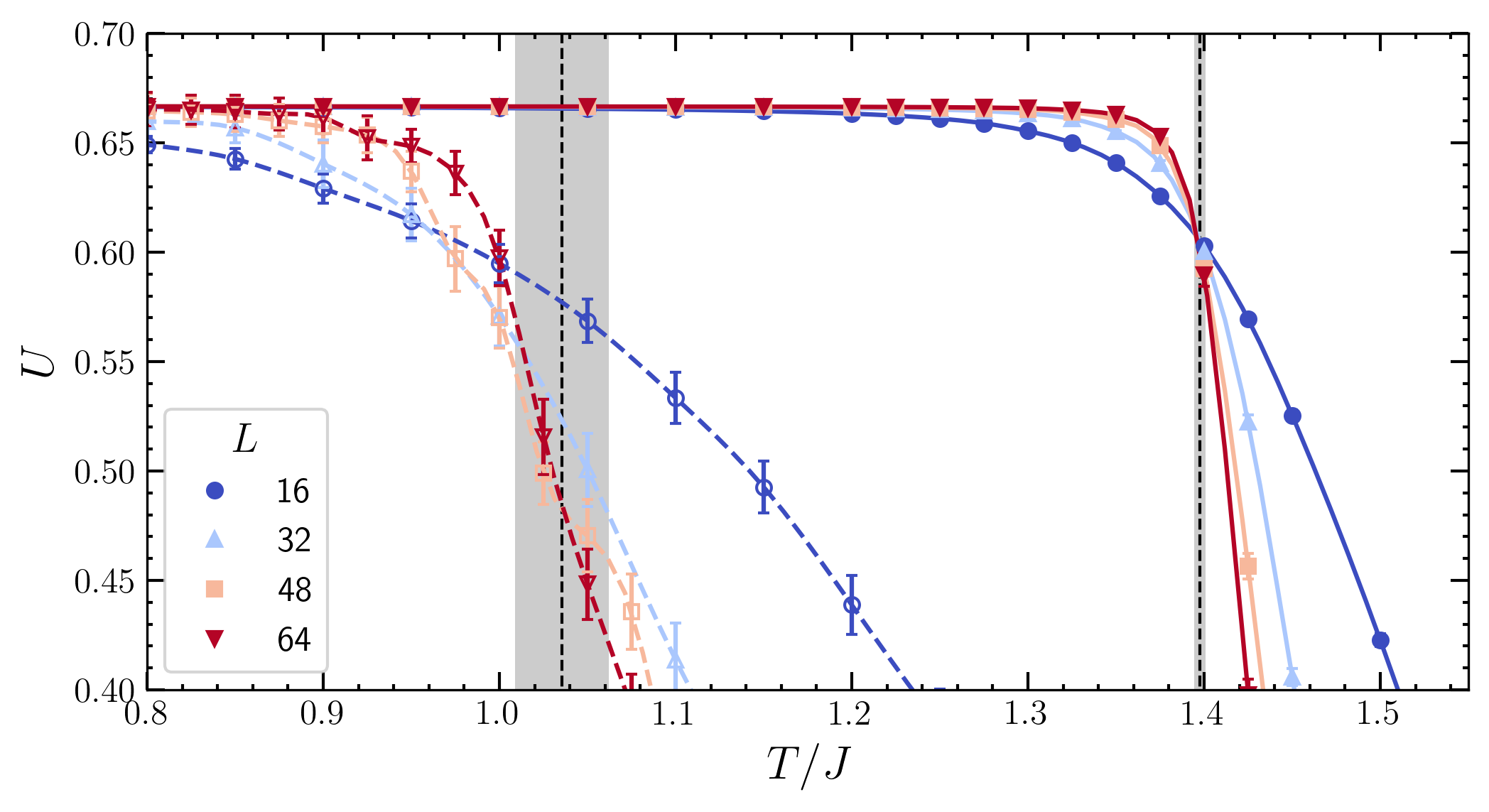}
    \caption{Binder ratio $U$ as a function of $T/J$ for different linear sizes $L$ at $\bar{D}/J=1$. Results are presented for $\sigma/J=0$ (solid lines, solid markers) and $\sigma/J=3$ (dashed lines, open markers). Black vertical lines mark the crossing between the two largest system sizes ($L=64,48$), and the shaded region is the distance to the next largest system crossing ($L=64,32$).}
    \label{fig:Binder_example}
\end{figure}

Finally, across the second order transition phase boundary, the critical temperature was obtained via the Binder ratio analysis. In Fig.~\ref{fig:Binder_example}, we present this quantity as a function of $T/J$ for $\sigma/J=0$ (solid lines) and $\sigma/J=3$ (dashed lines) for different system sizes. In the absence of disorder this quantity displays a sharp crossing at $T_c/J$, in excellent agreement with the literature~\cite{Beale1986,pittman2008,Silva2002,Kwak2015,Zierenberg2017,Fytas2019}. In the presence of large amounts of disorder ($\sigma/J=3$) the $L=16$ results are unreliable, but the results from larger system sizes $L\geq32$ provide a robust estimate of $T_c/J$.

\section{Local density approximation phase diagram}

\begin{figure}[tbp!]
    \centering
    \includegraphics[width=\linewidth]{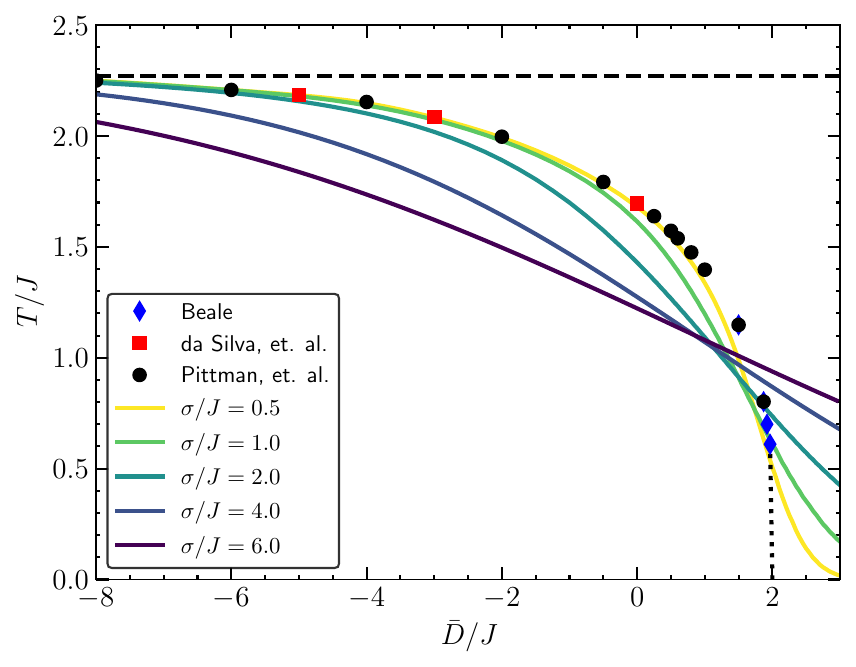}
    \caption{Phase diagram of the uniform Blume-Capel model as a function of $T/J$ and $D/J$. The symbols are data from representative manuscripts~\cite{Beale1986,Silva2002,pittman2008}. The lines are the approximate theory of Eq.~\eqref{eq:TC_LDA} for the indicated values of $\sigma/J$.} 
    \label{fig:UnifBCM}
\end{figure}

Although the LDA is unable to accurately reproduce the phase diagram, it provides useful insight into the $\sigma/J$ dependence of the model. For that reason, we present the predicted $T/J$ vs $\bar{D}/J$ phase diagram in Fig.~\ref{fig:UnifBCM} for different $\sigma/J$ within the LDA. These results predict that: (1) for $\bar{D}/J<2$, $T_c/J$ is suppressed, but remains finite as $\sigma/J$ increases, and (2) for $\bar{D}/J \geq2$, $T_c/J>0$, softening the first order transition.

\newpage 
\bibliography{Blume_Capel}

\end{document}